\documentclass[aps,prd,10pt,twocolumn, superscriptaddress,nofootinbib]{revtex4}
\usepackage{mathrsfs}  
\usepackage{epsfig, cancel}
\usepackage{latexsym}
\usepackage{natbib,comment}
\usepackage{mathrsfs,amssymb,amsmath,amsthm,amsfonts,tikz,graphicx,accents,hyperref, color}
\usepackage{url}
\usepackage{dcolumn}
\usepackage{multirow}
\usepackage{color}
\usepackage{cancel}
\usepackage{soul}
\usepackage[normalem]{ulem}
\usepackage{amsfonts,amssymb,amsmath, txfonts}
\usepackage{graphicx,epsfig}
\usepackage{psfrag}
\usepackage{subfigure}
\usepackage{hyperref}
\hypersetup{colorlinks=true}
\usepackage{mathtools}
\usepackage{enumitem}
\usepackage{float}
\usepackage{physics}
\usepackage{todonotes}
\usepackage[most]{tcolorbox}
\newcommand\eqbox[1]{\tcbhighmath{#1}}
\tcbset{highlight math style={left=02mm,right=02mm,top=02mm,bottom=02mm}} 

\def\H0{{\text{H}\hspace*{-2.05mm}\text{H} 0\hspace*{-1.35mm}0\ }}

\usepackage[all]{xy}
\usepackage{slashed}
\usepackage{slashed,ccaption}
\usepackage{multirow}
\usepackage{amsfonts}
\usepackage{amsmath}

\usepackage{caption}

\usepackage{array}
\hypersetup{ linktoc=all,
    colorlinks, linkcolor={palatinateblue},
    citecolor={brightpink}, urlcolor={amaranth}
}

\graphicspath{{Images/}}

\DeclareSymbolFont{extraup}{U}{zavm}{m}{n}
\DeclareMathSymbol{\varheart}{\mathalpha}{extraup}{86}
\DeclareMathSymbol{\vardiamond}{\mathalpha}{extraup}{87}
\makeatletter
\renewcommand*{\@fnsymbol}[1]{\ensuremath{\ifcase#1\or \clubsuit \or \vardiamond \or \varheart\or
    \spadesuit\or \mathparagraph\or \|\or **\or \dagger\dagger
    \or \ddagger\ddagger \else\@ctrerr\fi}}
\makeatother

\definecolor{rosy}{RGB}{230,235,252}
\definecolor{myframetitle}{RGB}{90,89,170}
\definecolor{myblocktitle}{RGB}{140,185,249}
\definecolor{mytitle}{RGB}{10,80,26}

\definecolor{darkgreen}{RGB}{27,130,45}
\definecolor{darkblue}{rgb}{0,0,0.3}
\definecolor{darkred}{rgb}{0.7,0,0}

\definecolor{light gray}{RGB}{220,220,220}
\definecolor{dark purple}{RGB}{108,0,217}
\definecolor{pink}{RGB}{190,20,100}
\definecolor{orang}{RGB}{193,63,0}
\definecolor{green}{RGB}{11,98,17}
\definecolor{darkpink}{RGB}{153,0,76}
\definecolor{bluegreen}{RGB}{0,102,102}
\definecolor{greenlagan}{RGB}{0,102,0}
\definecolor{redgreen}{RGB}{102,102,0}
\definecolor{Redgreen}{RGB}{153,76,0}
\definecolor{vividviolet}{rgb}{0.62, 0.0, 1.0}
\definecolor{amaranth}{rgb}{0.9, 0.17, 0.31}
\definecolor{palatinateblue}{rgb}{0.15, 0.23, 0.89}
\definecolor{brightpink}{rgb}{1.0, 0.0, 0.5}
\definecolor{cornflowerblue}{rgb}{0.39, 0.58, 0.93}
\definecolor{deepcarminepink}{rgb}{0.94, 0.19, 0.22}
\definecolor{radicalred}{rgb}{1.0, 0.21, 0.37}

\DeclareFontFamily{OT1}{rsfs}{}

\DeclareFontShape{OT1}{rsfs}{m}{n}{ <-7> rsfs5 <7-10> rsfs7 <10->rsfs10}{} 

\DeclareMathAlphabet{\mycal}{OT1}{rsfs}{m}{n}

\newcommand{\be}{\begin{equation}}
\newcommand{\ee}{\end{equation}}
\begin{document}

\title{Fluid/p-form duality}

\author{V.~Taghiloo}\email{v.taghiloo@iasbs.ac.ir}
\affiliation{School of Physics, Institute for Research in Fundamental
Sciences (IPM), P.O.Box 19395-5531, Tehran, Iran}
\affiliation{Department of Physics, Institute for Advanced Studies in Basic Sciences (IASBS),
P.O. Box 45137-66731, Zanjan, Iran}
\author{M.H.~Vahidinia}\email{vahidinia@iasbs.ac.ir}
\affiliation{Department of Physics, Institute for Advanced Studies in Basic Sciences (IASBS),
P.O. Box 45137-66731, Zanjan, Iran}
\affiliation{School of Physics, Institute for Research in Fundamental
Sciences (IPM), P.O.Box 19395-5531, Tehran, Iran}

\begin{abstract}
In this study, we demonstrate that an inviscid fluid in a near-equilibrium state, when viewed in the Lagrangian picture in $d+1$ spacetime dimensions, can be reformulated as a $(d-1)$-form gauge theory. We construct a fluid/$p$-form dictionary and show that volume-preserving diffeomorphisms on the fluid side manifest as a U(1) gauge symmetry on the {$(p+1)$-form} gauge theory side. {Intriguingly, Kelvin's circulation theorem and the mass continuity equation respectively appear as the Gauss law and the Bianchi identity on the gauge theory side.} Furthermore, we show that at the level of the sources, the vortices in the fluid side correspond to the $p$-branes in the gauge theory side. We also consider fluid mechanics in the presence of boundaries and examine the boundary symmetries and corresponding charges from both the fluid and gauge theory perspectives.
\end{abstract}
\maketitle

\section{Introduction}
A fluid is characterized as an assembly of particles that remain unchanged under relabeling. This relabeling is reflected at the formulation level as invariance under volume-preserving diffeomorphisms. One of the direct implications of these diffeomorphisms is the conservation of circulation, commonly referred to as the Kelvin circulation theorem \cite{RevModPhys.70.467,1990AdGeo..32..287S, Salmon1988HAMILTONIANFM, V.Arnold}.

The examination of a fluid in a state close to equilibrium simplifies the analysis. This is because the fluctuations around this equilibrium state at the leading order are effectively described through a linear theory. Although volume-preserving diffeomorphisms form an infinite-dimensional non-Abelian algebra, they become Abelian at the linearized level. This suggests that the near-equilibrium state of the fluid could be described using an Abelian gauge theory. Intriguingly, we demonstrate in this paper that this is indeed the case. 

{To identify the gauge theory that describes the fluid near-equilibrium state, we note that the global charges in gauge theories are represented by lower-dimensional integrals, which is a distinguishing feature of gauge theories.}
 On the other hand, in the fluid context, the Kelvin circulation theorem asserts that the line integral of the fluid velocity (circulation) is conserved, a statement that holds in any spacetime dimension. Therefore, we anticipate that the corresponding dual gauge theory should possess a global charge expressed by a one-dimensional integral. Simultaneously, we understand that a $(p+1)$-form gauge theory in $(d+1)$-spacetime dimension results in a global charge that is a $d-p-1$ dimensional integral \cite{Henneaux:1986ht}. This leads us to infer that the corresponding gauge theory could be a $(d-1)$-form gauge theory. {In this paper, we confirm the Lagrangian description of inviscid fluid in $(d+1)$-spacetime dimensions, can indeed be reinterpreted as a $(d-1)$-form gauge theory, as long as the fluid remains in the near-equilibrium state.}

We will establish a dictionary that relates the fluid variables to the gauge theory quantities. According to this dictionary, the fluid velocity and density correspond to the electric and magnetic fields on the gauge theory side, respectively. {Interestingly, in this context, the Kelvin circulation theorem and the mass continuity equation serve as the counterparts to the Gauss law and the Bianchi identity, respectively.}  {Comparably,} the Gauss law ensures the gauge invariance in gauge theory, and the Kelvin circulation theorem  {guarantees} the invariance of the fluid under volume-preserving diffeomorphisms. Furthermore, the dictionary goes a step further and relates the sources of equations in both fluid and gauge theory sides. Intriguingly, $(d-2)$-branes \footnote{The term ``brane" comes from ``membrane" which refers to the two-dimensional object.}, which serve as the source of the $(d-1)$-form gauge theory, are found to be dual to vortices on the fluid side. {A comprehensive summary of our fluid/gauge dictionary can be found in Table \ref{dictionary}.}

{As previously mentioned,} we provide a dual gauge theory description for an inviscid fluid within the Lagrangian framework. Recently, the study of topological characteristics in fluid dynamics has prompted the exploration of potential gauge theory descriptions for fluid systems (see for example \cite{Delplace_2017, Venaille_2021}).  All these gauge theory formulations are presented in the Eulerian picture. For instance, a gauge theory description of shallow water was initially introduced in \cite{Tong:2022gpg}, with its implications further explored in \cite{Sheikh-Jabbari:2023eba}. Gauge theory descriptions for incompressible $1+2$ and $1+3$ dimensional Eulerian fluids were examined in \cite{Eling:2023iyx, Eling:2023apf}. Additionally, a gauge theory description for a compressible $1+2$ dimensional Eulerian fluid was also discussed in \cite{Nastase:2023rou}. Also, $1+2$ dimensional fluids were considered through master action in \cite{Dayi:2023ckd}.

The structure of this paper is as follows: In Section \ref{sec:fluid}, we provide a description of an inviscid fluid using the Lagrangian approach and establish our notation. Section \ref{sec:pform} introduces the dual gauge theory of the Lagrangian fluid, along with its associated symmetries and charges. In Section \ref{More-Lagrangian-Fluid}, we delve into the symmetries of the Lagrangian fluid, examining them through the lens of fluid variables.

\section{Fluid in Lagrangian picture}\label{sec:fluid}
In this section, we examine an inviscid fluid in the Lagrangian description and construct an action that is invariant under volume-preserving diffeomorphisms. This action builds upon the one presented in \cite{Bahcall:1991an, Susskind:2001fb}, extending it from its original formulation in $2+1$ spacetime dimensions to more general $d+1$ spacetime dimensions.

Let us start with a group of non-relativistic particles \footnote{By particle we mean a parcel of fluid at mesoscopic scale which allows a classical description.}, all having the same mass $m$, in a $d+1$ spacetime dimension. The coordinates of these particles are represented by $\vb{x}_{\alpha}=\{x^{i}_{\alpha}\}$, where $\alpha=1,\cdots, N$ denotes the discrete particle label and $i=1,\dots,d$ shows the space coordinates.
The dynamics of these particles are described using the following  Lagrangian
\begin{equation}\label{action-discrete}
    \text{L}=\sum_{\alpha} \frac{1}{2}m \dot{\vb{x}}_{\alpha}^2-\int \dd^d x\, U(\vb{x},\vb{x}_{\alpha},\dot{\vb{x}}_{\alpha})\, .
\end{equation}
The potential term arises due to the interactions between particles and the existence of external sources. 

We focus on the continuum limit, where we replace discrete labels with a continuous variable, denoted by 
$\vb{x}_\alpha(t) \to \vb{x}(\vb{y},t)$. In this case, $\vb{y}$ represents the continuous label of particles. The $y$ coordinate can be thought of as representing the initial position of fluid particles $\vb{y}=\vb{x}(\vb{y},t_0)$. 

The density of particles in $y$ and $x$ spaces are denoted by $\rho_y$ and $\rho_x$, respectively. Without loss of generality, we assume that the density in the $y$-space is constant and equal to $\rho_0$, denoted by $\rho_0:=\rho_y$ \footnote{
To demonstrate that this assumption is always valid, we can consider the relationship between the fluid densities in two different $y$-spaces, given by $\rho_y'=\left|\frac{\partial y}{\partial y'}\right|\rho_y$. If $\rho_y$ is not constant, we can perform a relabeling such that $\rho_y'$ becomes constant. It is important to note that this type of relabeling is not unique. In fact, there are an infinite number of ways to perform such a relabeling.}. {In addition, for simplicity, we will employ the notation $\rho:=\rho_x$ hereafter.} The relationship between the real space density, $\rho$, and the $y$-space, $\rho_0$, is given by
\begin{equation}\label{eq:real-density}
{\rho}=\left|\frac{\partial \vb{y}}{\partial \vb{x}}\right|\rho_0\, ,
\end{equation}
where $\left|\frac{\partial \vb{y}}{\partial \vb{x}}\right|$ represents the Jacobian matrix of the transformation between the $y$ and $x$ coordinates.
The latter is simply the mass continuity equation
\begin{equation}\label{continuity-eq}
    \partial_{t}\rho+\rho_{0}\partial_{i}\dot{x}_{i}=0\, .
\end{equation}
To get this one should take
$\rho(\vb{y},t_0+\delta t)$ and $\vb{x}(\vb{y},t_0+\delta t)$ in \eqref{eq:real-density}.

As we will demonstrate in the following, the continuum version of \eqref{action-discrete} is given as
\begin{equation}\label{action-1}
    S[\vb{x}(\vb{y},t)]=\int d t\int d^{d} y\, \rho_0 \left[\frac{m}{2}\dot{\vb{x}}^2-V\left(\vb{x},\dot{\vb{x}},\rho_0 \abs{\pdv{\vb{y}}{\vb{x}}}\right)\right]\,.
\end{equation}
It is crucial to note that this action is invariant under volume-preserving diffeomorphisms on $y$ space
\begin{equation}
\vb{y}\to \vb{y}'(\vb{y}) \quad \text{such that} \quad \abs{\pdv{\vb{y}'}{\vb{y}}}=1\, ,
\end{equation}
where $\vb{x}'(\vb{y}',t)=\vb{x}(\vb{y},t)$ is assumed. Note that fluids are distinguished from solids by their enhanced symmetry, namely, volume-preserving diffeomorphisms \cite{RevModPhys.70.467,1990AdGeo..32..287S, Salmon1988HAMILTONIANFM, V.Arnold}.

Now to obtain action \eqref{action-1}, we assume that the interparticle interactions are short-range. This allows us to consider the potential $U$ as a function of the first derivatives of $\vb{x}$ namely, 
$U\qty(\vb{x},\dot{\vb{x}},\pdv{\vb{x}}{\vb{y}})$. In addition, to maintain the volume-preserving diffeomorphisms, which is essential to define fluid, we must assume that a specific combination of first-order derivatives $\partial x_i/\partial y_j$, appears in the action. The only possible combination that preserves this symmetry is $\left|\frac{\partial \vb{y}}{\partial \vb{x}}\right|$.  We also note that the relation between $U$ and $V$ is given by $V=U/\rho$. This difference is a result of converting the integral from real space to $y$-space in \eqref{action-discrete} \footnote{We note that in general, the potential can be dependent on the temperature. {As far as the near equilibrium state is concerned the temperature remains constant. In this case,  the velocity of particles depends on $T$ implicitly.} Under the assumption that we are dealing with zero temperature, we will proceed accordingly to circumvent this complexity.}.

We conclude this part by emphasizing that our consideration is limited to the volume-preserving diffeomorphisms in the $y$-space, not the $x$-space. This is because we use the Lagrangian picture to describe fluid dynamics. The $x$-space diffeomorphisms are closely tied to the Eulerian picture, which we do not explore in this manuscript.
\subsection{Volume-preserving diffeomorphisms}\label{sec:fluid-diffeos}
By construction, the action \eqref{action-1} is invariant with respect to volume-preserving diffeomorphisms in $y$-space. In this line of argument, we treat $x_{i}(y,t)$ as a collection of scalar fields with respect to these diffeomorphisms, $x'_i(y',t)=x_i(y,t)$. In the following, we will focus solely on the infinitesimal version of these diffeomorphisms
\begin{equation}\label{inf-diffeos}
    y'_{i}=y_i+\mathcal{Y}_{i}(y)\, .
\end{equation}
In terms of infinitesimal diffeomorphisms, the condition for preserving volume is expressed as $\frac{\partial\mathcal{Y}_{i}}{\partial y^{i}}=0$. The Poincare lemma indicates one can solve this condition locally as follows
\begin{equation}\label{volume-pres-cond-2}
\mathcal{Y}^i=\partial_{j}\lambda^{ij}\, \hspace{.25 cm} \text{with} \hspace{.25 cm} \lambda^{ij}=-\lambda^{ji}\, ,
\end{equation}
where for the sake of convenience,  the shorthand notation $\partial_{i}:=\frac{\partial}{\partial y_i}$ is used.
It is worth noting that these volume-preserving diffeomorphisms form a non-abelian algebra with the standard Lie bracket
\begin{equation}\label{sym-algebra}
[\mathcal{Y},\tilde{\mathcal{Y}}]^i=\mathcal{Y}^{j}\partial_{j}\tilde{\mathcal{Y}}^i-\tilde{\mathcal{Y}}^{j}\partial_{j}\mathcal{Y}^i\, .
\end{equation}
The infinitesimal variation of the $x$-field under this transformation is given by
\begin{equation}\label{diff-inv}
    \delta_{\mathcal{Y}} x_i=\mathcal{L}_{\mathcal{Y}}x_i=\mathcal{Y}^{j}{\partial_j x_i}\, ,
\end{equation}
where $\mathcal{L}_{\mathcal{Y}}$ denotes the Lie derivative along $\mathcal{Y}^i$. In the following section, we provide a gauge theoric description of this fluid where the mentioned diffeomorphism plays the role of gauge symmetry. 
\section{Gauge theory description}\label{sec:pform}
In this section, we examine the near-equilibrium state of a fluid and demonstrate that it can be effectively described using a $(d-1)$-form gauge theory. {Remarkably,} our analysis reveals a correspondence between volume-preserving diffeomorphisms in the fluid and $\text{U(1)}$ gauge symmetries in the gauge theory.
\subsection{Volume preserving diffeomorphisms as gauge symmetries}\label{sec:gauge-symmetry}
We are interested in considering the near-equilibrium state of the fluid. In this regard, we assume the fluid potential has a minimum at $\vb{x}=\vb{y}$, $\dot{\vb{x}}=0$, and $\rho=\rho_0$, namely,
\begin{equation}
    \frac{\partial V}{\partial x_i}\big|_{\text{eq.}}=0\, , \hspace{1 cm} \frac{\partial V}{\partial \dot{x}_i}\big|_{\text{eq.}}=0\, , \hspace{1 cm} \frac{\partial V}{\partial \rho}\big|_{\text{eq.}}=0\, .
\end{equation}
 This equilibrium condition implies that $x_i(\vb{y},t)=y_i$ is a solution to equations of motion (see \eqref{eom-fluid} for more details), which represents the equilibrium state. We can now examine small deviations from this equilibrium 
\begin{equation}\label{fluctuation}
    x_i=y_i+ \mathcal{A}_{i}(y)\,.
\end{equation}
According to this equation, the volume-preserving diffeomorphisms \eqref{diff-inv} act on $\mathcal{A}_i$ as follows
\begin{equation}
    \delta_{\lambda} \mathcal{A}_{i}= \delta_{\lambda} x_{i}={\frac{\partial x_i}{\partial y_j}}\partial_{k}\lambda^{jk}\, ,
\end{equation}
where  \eqref{volume-pres-cond-2} is used. By substituting \eqref{fluctuation} in the above expression, we get
\begin{equation}
    \delta_{\lambda} \mathcal{A}_{i}=\partial_{j}\lambda^{ij}+\partial_{j}\mathcal{A}_{i}\partial_{k}\lambda^{jk}\, .
\end{equation}
We neglect the second nonlinear term  in the above equation and restrict ourselves to the linear transformation
\begin{equation}\label{gauge-trans-1}
    \delta_{\lambda} \mathcal{A}_{i}=\partial_{j}\lambda^{ij}\, .
\end{equation}
Interestingly, this equation is reminiscent of the gauge transformation for the gauge field $\mathcal{A}$. In this sense, the volume-preserving diffeomorphisms are related to gauge symmetries. In the following, we concrete this statement and argue how the dynamics of the Hodge dual of $\mathcal{A}$ is described by a $(d-1)$-form gauge theory. 
\subsection{$(d-1)$-form theory as an effective theory of near-equilibrium state}\label{sec:gauge-symmetry}
In this subsection, we construct the effective action describing the near-equilibrium state of the fluid. To simplify the analysis, we further assume that the potential is solely a function of density, represented as $V=V(\rho)$. In this regard, we expand the action \eqref{action-1} around the equilibrium state. We start with the expansion of the real space density \eqref{eq:real-density}
\begin{equation}\label{eq:rhoexp}
    \rho\approx\rho_0(1-\partial_{i}\mathcal{A}_i)\, .
\end{equation}
Now we are ready to expand the potential energy
\begin{equation}
    V(\rho)=V_0+a\,(\partial_{i}\mathcal{A}_i)^2+\mathcal{O}(\mathcal{A}^3)\, ,
\end{equation}
where $ V_{0}=V\big|_{\text{eq.}}$ and $a=\frac{1}{2}\rho_0^2 \frac{\partial^2 V}{\partial \rho^2}\Big|_{\text{eq.}}$.
Hence the Lagrangian up to quadratic order in $\mathcal{A}_i$, gives
\begin{equation}
   \rho_0^{-1}\mathcal{L}=\frac{m}{2}\dot{\mathcal{A}}_{i}^2-a(\partial_{i}\mathcal{A}_{i})^2\, .
\end{equation}
To write this Lagrangian in a more familiar form, let us introduce $(d-1)$-form $A_{i_1\cdots i_{d-1}}$ as the Hodge dual of $\mathcal{A}_{i}$
\begin{equation}\label{eq:HodgeDual}
\mathcal{A}_{i}=\frac{1}{(d-1)!}\epsilon_{i_1 \cdots i_{d-1} i}\, A^{i_1 \cdots i_{d-1}}\, ,
\end{equation}
and hence
\begin{equation}
    \begin{split}\label{Lag-pform-1}
        \rho_0^{-1}\mathcal{L}&=\frac{m}{2}\frac{1}{(d-1)!}\dot{A}_{i_1\cdots i_{d-1}}^2-a \frac{d}{(d-1)!}\left(\partial_{[i_1}A_{i_2\cdots i_{d}]}\right)^2\, .
    \end{split}
\end{equation}
It suggests to define an effective velocity (sound  velocity of the fluid) as
\begin{equation}
    c^2:={\frac{2a}{m}}\, . 
\end{equation}
Now we introduce the $d$-form field strength as follows
\begin{equation}
    \bold{F}=d\bold{A}\,, \hspace{1 cm}  F_{\mu_1 \cdots \mu_d}=d\, \partial_{[\mu_1}A_{\mu_2 \cdots \mu_d]}\, .
\end{equation}
In the temporal gauge $A_{0 i_1 \cdots i_{d-2}}=0$, we have
\begin{equation}
    F_{0 i_2 \cdots i_d}=\partial_{t}{A}_{i_2 \cdots i_d}\, , \hspace{.7 cm} F_{i_1 \cdots i_d}=d\, \partial_{[i_1}A_{i_2 \cdots i_d]}\, .
\end{equation}
Finally, the effective action characterizing the near-equilibrium state can be reduced to the following well-known form
\begin{equation}\label{p-from-action}
\eqbox{S=-\frac{1}{2g^2}\int \dd^{d+1}y\,  F_{\mu_1 \cdots \mu_d}F^{\mu_1 \cdots \mu_d}\, ,}
\end{equation}
where $g^2:={\frac{d!}{2a\rho_0}}$.
This is the seminal action of a $(d-1)$-form gauge theory.

Note that, the Lagrangian \eqref{Lag-pform-1} (or equivalently action \eqref{p-from-action} in the temporal gauge) is invariant under the following gauge transformation $ \delta_{\Lambda} A_{i_1 \cdots i_{d-1}}=(d-1)\partial_{[i_1}\Lambda_{i_2\cdots i_{d-1}]}$, where $\Lambda$ is an arbitrary time independent $(d-2)$-form.
By using the  $\Lambda_{i_1 \cdots i_{d-2}}:=\frac{1}{2!}\epsilon_{i_1 \cdots i_{d-2}ij}\lambda^{ij}$, the latter is nothing else than \eqref{gauge-trans-1}. This is the standard form of a gauge transformation for a gauge field of rank $(d-1)$. For the case where $d=2$, this result is consistent with the findings of Bahcall and Susskind \cite{Bahcall:1991an, Susskind:2001fb}.

One may note that \eqref{p-from-action} is invariant under more general gauge transformation
 \begin{equation}
   \eqbox{\delta_{\Lambda} A_{\mu_1 \cdots \mu_{d-1}}=(d-1)\partial_{[\mu_1}\Lambda_{\mu_2\cdots \mu_{d-1}]}\, .}
\end{equation}
However, this gauge transformation needs to be consistent with the near-equilibrium condition \eqref{fluctuation} (see \cite{Susskind:2001fb}). In the following section, we will consider the conserved charge associated with these symmetries and demonstrate that the global part of this gauge transformation is nothing more than the Kelvin circulation.
{We now conclude this subsection by providing a summary of our dictionary
\begin{equation}\label{guage-fluid-dictionary}
    \eqbox{\begin{split}
        &\dot{x}_{i_1}=\frac{(-1)^{d}}{(d-1)!}\epsilon_{i_1 i_2 \cdots i_d} F^{0 i_2 \cdots i_d}\, , \\
        &\frac{\rho}{\rho_0}=1+\frac{(-1)^{d}}{d!}\epsilon_{i_1 i_2 \cdots i_d} F^{i_1 \cdots i_d}\, ,
    \end{split}}
\end{equation}
where equations \eqref{fluctuation}, \eqref{eq:rhoexp}, and \eqref{eq:HodgeDual} are used. {Hence, 
the electric part of the gauge field gives the Lagrangian velocity of the fluid while the magnetic part is related to its mass density fluctuations around the equilibrium.}
Presenting the dictionary in terms of the velocity field, as opposed to the fluid’s particle coordinates, offers the advantage of gauge invariance. This means it is applicable across all gauges, not just the temporal gauge.
\subsection{Symmetry and conserved charges}\label{sec:symmetry-charge}
In this subsection, we will examine the symmetries and corresponding charges of the near-equilibrium fluid on the dual gauge theory side. 
The symmetries of $(p+1)$-form gauge theories have been studied in \cite{Afshar:2018apx, Compere:2007vx, Esmaeili:2020eua}, in what follows we only investigate the Noether charge and its interpretation for $p=d-2$.

As a starting point, we will utilize the form field action \eqref{p-from-action}, expressed in the standard index-free differential form notation
\begin{equation}
   S[\bold{A}]=-\frac{1}{2 g^2}\int   \star \bold{F}\wedge\bold{F}\, .
\end{equation}
Clearly, this action is invariant with respect to the following U(1) gauge transformations
\begin{equation}\label{gauge-transformation-A}
    \bold{A}\to \bold{A}+\dd\bold{\Lambda}\, , \hspace{1 cm}  \delta_{\bold{\Lambda}}\bold{A}=\dd\bold{\Lambda}\, ,
\end{equation}
where $\bold{\Lambda}$ is an arbitrary $(d-2)$-from. A generic variation of the action together integration by parts yields
\begin{equation}
    \delta S[\bold{A}]=-\frac{1}{g^2}\int (\mathcal{E}\wedge \delta \bold{A}+\dd\bold{\Theta})\, ,
\end{equation}
where $\mathcal{E}$ and $ \bold{\Theta}$ are respectively the equation of motion and the symplectic potential
\begin{equation}\label{EOM-symp-pot}
    \mathcal{E}:=\dd(  \star \bold{F})=0\, , \hspace{1 cm} \bold{\Theta}[\bold{A},\delta\bold{A}]:=-  \star \bold{F}\wedge \delta \bold{A}\, .
\end{equation}
The equations of motion nothing else than $\partial_{\mu_1}F^{\mu_1 \cdots \mu_d}=0$. One can decompose this equation into two parts
\begin{subequations}
    \begin{align}
        &\partial_{i_1}F^{0i_1\cdots i_{d-1}}=0\, ,\label{Gauss-law}\\
        &\partial_{t}F^{t i_2 \cdots i_d}+\partial_{i_1}F^{i_1 i_2 \cdots i_d}=0\, .
    \end{align}
\end{subequations}
The first equation \eqref{Gauss-law} reminds us of the standard Gauss law (or Gauss constraint). In the temporal gauge, by noting  \eqref{fluctuation}, one can rewrite it in terms of fluid variables as follows
\begin{equation}
    \epsilon^{i_1 \cdots i_d}\partial_{i_1}\dot{x}_{i_2}=0\, .
\end{equation}
Hence the Gauss law in the fluid language states that the fluid is irrotational. For example, in $1+3$ dimensional spacetime, it yields $\nabla\times \dot{\vb{x}}=0$. 

A natural generalization of the homogeneous equations of motion \eqref{EOM-symp-pot} is given by adding the source on the right-hand side, namely,
\begin{equation}\label{eom-J}
    \dd(  \star \bold{F})= \star\bold{J}\, .
\end{equation}
In this context, $\bold{J}$ represents a $(d-1)$-form and serves as an external source in $(d-1)$-form gauge theories. On the fluid side, adding a non-zero term to the right-hand side is equivalent to introducing a vortex into the fluid system. This demonstrates that external charged objects in the $(d-1)$-form gauge theories are dual to vortices in the fluid. Charged objects in $(p+1)$-form gauge theories are known as $p$-branes, and in our case, the extended charged particles are $(d-2)$-branes. For instance, in a $1+2$ dimensional spacetime, they are point particles, while in a $1+3$ dimensional spacetime, they are strings. On the fluid side, these are equivalent to vortices. We refer to the duality between the source and extended charged objects as \textit{brane/vortex duality}.

To further illustrate the connection of sources on two sides, we note that as a consequence of the integrability \eqref{eom-J} we have
\begin{equation}
    \dd  \star\bold{J}=0\, .
\end{equation}
On the gauge theory side, this represents the conservation of the external current. Interestingly, on the fluid side, this results in the \textit{Kelvin circulation theorem}. In an inviscid fluid, the vorticity is locally conserved and its integration over a surface element gives the circulation which is a conserved quantity. As we will show subsequently, this conservation law is a consequence of the Noether theorem in the gauge theory description of the fluid.

{We close this subsection by noting that the Bianchi identity on the gauge theory side corresponds to the mass continuity equation on the fluid side. To show this, we note that the Bianchi identity, $\star\dd \bold{F}=0$, is given by
\begin{equation}\label{Bianchi}
    \epsilon^{i_1 \cdots i_d}\left(\partial_{t}F_{i_1 \cdots i_d}-d \partial_{i_1}F_{0 i_1 \cdots i_d}\right)=0\, .
\end{equation}
Now by using the fluid/gauge dictionary \eqref{guage-fluid-dictionary}, we find this equation is nothing else than the linearized mass continuity equation \eqref{continuity-eq}.
}
\subsection{Noether current and Noether charge} 
Now, we consider the gauge symmetries \eqref{gauge-transformation-A} in the context of the Noether theorem. Given an action and a set of symmetry transformations as input, the Noether systematic approach produces the corresponding conserved charges as output. To accomplish this, we begin with the Noether $d$-form current associated with these symmetries
\begin{equation}\label{Noether-current-1}
    \bold{J}_{{\Lambda}}:=\bold{\Theta}[\bold{A},\delta_{{\Lambda}}\bold{A}]=-  \star \bold{F}\wedge \delta_{\Lambda} \bold{A}=- \star\bold{F}\wedge \dd\bold{\Lambda}\, .
\end{equation}
First of all, we consider the conservation of this Noether current
\begin{equation}
    \dd\bold{J}_{{\Lambda}}=-\dd  \star \bold{F}\wedge \dd\bold{\Lambda}+ \star \bold{F}\wedge \dd^{2}\bold{\Lambda}\, .
\end{equation}
Now by virtue of the identity $\dd^2=0$, we find
\begin{equation}
    \dd\bold{J}_{{\Lambda}}=-\mathcal{E}\wedge \dd\bold{\Lambda}\, ,
\end{equation}
here we used \eqref{EOM-symp-pot}. This result shows the Noether current is conserved on-shell $ \dd\bold{J}_{{\Lambda}}=0$.

Due to the gauge nature of the underlying symmetry, we anticipate that one can express the Noether current as an exact form. To do so, we write the Noether current as follows
\begin{equation}
    \bold{J}_{{\Lambda}}=- \star \bold{F}\wedge \dd\bold{\Lambda}=-\mathcal{E}\wedge \bold{\Lambda}+\dd(  \star \bold{F}\wedge \bold{\Lambda})\, .
\end{equation}
Then by using the on-shell condition, $\mathcal{E}=0$, we obtain the low dimensional version of the Noether current
\begin{equation}\label{codim-2-charge}
    \bold{J}_{{\Lambda}}=\dd\bold{Q}_{\Lambda}\, , \hspace{1 cm} \bold{Q}_{\Lambda}:= \star \bold{F}\wedge \bold{\Lambda}\, .
\end{equation}
Now we are ready to simply define the Noether charge associated with the above Noether current 
\begin{equation}\label{codim-1-charge-p-form}
    Q_{\Lambda}:=\int_{\Sigma_d} \bold{J}_{{\Lambda}}=\oint_{\Sigma_{d-1}} \bold{Q}_{\Lambda}\, ,
\end{equation}
where $\Sigma_{d}$ is a $d$-dimensional Cauchy surface {(a time-constant hypersurface)} and $\Sigma_{d-1}$ is its $(d-1)$-dimensional boundary.
{The last equality was achieved by using the Stokes theorem.} The explicit form of this codimension two charge (surface charge) is given by
\begin{equation}\label{Surface-Charge}
   \eqbox{  Q_{\Lambda}[\Sigma_{d-1}]=\oint_{\Sigma_{d-1}}   \star \bold{F}\wedge \bold{\Lambda}\, .}
\end{equation}
For any choice of gauge parameters $\bold{\Lambda}$, $Q_{\Lambda}$ is conserved. However, there exist certain distinguished gauge transformations that keep $\bold{A}$ invariant, $\delta_{\bold{\Lambda}}\bold{A}=0$. The corresponding charges of these $\bold{\Lambda}$'s are referred to as global charges. Here is an easy-to-understand explanation that shows circulation is a certain global charge.
To show that, we assume $\bold{\Lambda}$ is proportional to volume form of a $d-2$ surface
\begin{equation}
\bold{\Lambda}=(-1)^{d-1}\delta^{d-2}(\Sigma_{d-2})\boldsymbol{\epsilon}_{{d-2}}.
\end{equation}
In this equation, {codimension-two surface} $\Sigma_{d-1}$ is defined as the product of $\Sigma_{d-2}$ and a closed curve $\mathcal{C}$, namely $\Sigma_{d-1}:=\Sigma_{d-2}\times \mathcal{C}$.
{The $q$-form $\boldsymbol{\epsilon}_q$ represents the normalized volume form on surface $\Sigma_{q}$}
\footnote{Its explicit form is given by $\boldsymbol{\epsilon}_{q}:=\frac{1}{q!\, V(\Sigma_{q})}\epsilon_{i_{1}\cdots i_{q}}\dd{y}^{i_1}\wedge \cdots \wedge \dd{y}^{i_{q}}$ where $V(\Sigma_{q})$ denotes the volume of $\Sigma_{q}$.}.
It is straightforward to verify that $\delta_{\bold{\Lambda}}\bold{A}=\dd\bold{\Lambda}=0$. By using this choice, equation \eqref{Surface-Charge} simplifies to a line integral
\begin{equation}\label{eq:Q-p-Kelvin}
  \eqbox{  \Gamma_{\mathcal{C}}=(-1)^{d-1}\oint_{\mathcal{C}}   \star\bold{F}\, ,}
\end{equation}
where $ \Gamma_{\mathcal{C}}:=Q[\mathcal{C}]$. Now let us rewrite this charge in terms of the fluid variables by noting the duality between velocity and field strength \eqref{guage-fluid-dictionary}
\begin{equation}
    \Gamma_{\mathcal{C}}=\oint_{\mathcal{C}}\, \dot{\vb{x}}\cdot \dd\vb{\ell}\, .
\end{equation}
This is nothing else than the Kelvin circulation theorem for a dissipation-less fluid in the Lagrangian picture. In fluid mechanics, the Kelvin circulation theorem states that the circulation associated with an arbitrary closed curve $\mathcal{C}$, denoted by $\Gamma_{\mathcal{C}}$, is conserved. This result is a consequence of volume-preserving diffeomorphism.
Consequently, this result offers us a reinterpretation of the Kelvin circulation theorem through the lens of Noether’s theorem and conserved charges. As a final point, it’s worth noting that the surface charges, as defined in \eqref{Surface-Charge}, form an Abelian U(1) algebra.

We conclude this subsection with a discussion on the magnetic charge. We begin by defining the following $d$-form magnetic conserved current
\begin{equation}\label{cons-current-n-m}
    \bold{J}_{d}=\bold{F}\, , \hspace{1 cm} \dd\bold{F}=0\, .
\end{equation}
From this, we  derive the following conserved magnetic charge
\begin{equation}
    Q_m=\int_{\Sigma_d} \bold{F}=\oint_{\mathcal{S}_{d-1}} \bold{A}\, .
\end{equation}
In terms of fluid variables, it yields
\begin{equation} 
    Q_m=\oint_{\mathcal{S}_{d-1}} \dd^{d-1}s\, \vb{n}\cdot(\vb{x}-\vb{y})\, ,
\end{equation}
where $\vb{n}$ is the normal vector on surface $\mathcal{S}_{d-1}$. This result indicates that the sum of fluid fluctuations around equilibrium over the codimension-two surface $\mathcal{S}_{d-1}$ remains conserved.

\subsection{Towers of conserved charges}
Taking inspiration from equation \eqref{Noether-current-1}, we may define 
the following tower of conserved currents
\begin{equation}\label{cons-current-n}
    \bold{J}_{n+2}=-  \star \bold{F}\wedge \dd \bold{\Lambda}_{n}\, , \hspace{1 cm} \forall 0\leq n \leq d-1\, .
\end{equation}
Here indices indicate the rank of the form fields. These currents are conserved on-shell, namely,
\begin{equation}
    \dd\bold{J}_{n+2}=-\dd  \star \bold{F}\wedge \dd \bold{\Lambda}_{n}+  \star \bold{F}\wedge \dd^2 \bold{\Lambda}_{n}
\end{equation}
The first term vanishes by applying the equation of motion, $\dd  \star \bold{F}=0$, and the second term vanishes through the identity $\dd^2=0$. By applying the equation of motion, one can rewrite the conserved currents \eqref{cons-current-n} as follows
\begin{equation}
    \bold{J}_{n+2}=\dd \bold{Q}_{n+1}\, , \hspace{1 cm} \bold{Q}_{n+1}:=  \star \bold{F}\wedge \bold{\Lambda}_{n}\, .
\end{equation}
From this, we  define the following tower of conserved charges
\begin{equation}
    Q=\int_{\Sigma_{n+2}} \bold{J}_{n+2}=\oint_{\mathcal{S}_{n+1}}\bold{Q}_{n+1}\, ,
\end{equation}
where $\mathcal{S}_{n+1}:=\partial \Sigma_{n+2}$. The explicit form of these charges is
\begin{equation}\label{Tower-charges}
Q[\bold{\Lambda}_{n};\mathcal{S}_{n+1}]=\oint_{\mathcal{S}_{n+1}}  \star \bold{F}\wedge \bold{\Lambda}_{n}\, , \hspace{.7 cm} \forall 0\leq n \leq d-1\, ,
\end{equation}
{where $n=0$ case for constant $\Lambda$ is proportional to Kelvin circulation $\Gamma_{C}$.}
 These tower of charges \eqref{Tower-charges} form a U(1) Abelian Lie algebra as follows
\begin{equation}
    \Big\{Q[\bold{\Lambda}_{d-n-1};\mathcal{S}_{d-n}],Q[\bold{\Lambda}'_{d-m-1};\mathcal{S}_{d-m}]\Big\}=0\,.
\end{equation}
One can interpret the tower of charges \eqref{Tower-charges} as the charges of higher form symmetries which are symmetries whose charged objects are extended operators supported on lines, surfaces, etc. (for a review see \cite{Gomes:2023ahz, Cordova:2022ruw, Schafer-Nameki:2023jdn, Brennan:2023mmt, Sharpe:2015mja, McGreevy:2022oyu}). To our knowledge, this is the first time that the higher form symmetries appear in the context of fluid mechanics.

{Remarkably, it is possible to understand the tower of charges \eqref{Tower-charges} in the framework of the Noether theorem as follows:
Let us start from \eqref{Surface-Charge} and take  $\Sigma_{d-2}=\Sigma_{n}\times \Sigma_{d-n-2}$ and 
\begin{equation} 
\bold{\Lambda}=\bold{\Lambda}_{n}\wedge \bold{\Lambda}_{d-n-2}\,.
\end{equation}
Then by choosing $\bold{\Lambda}_{d-n-2}=\delta^{d-n-2}(\Sigma_{d-n-2})\boldsymbol{\epsilon}_{d-n-2}$
we obtain \eqref{Tower-charges} as a certain Noether charge
\footnote{We are thankful to M.M. Sheikh-Jabbari for pointing this out.}. Interestingly, in this way the higher form charges \eqref{Tower-charges} arise from the standard Noether theorem.
}

\section{More on Lagrangian Fluid}\label{More-Lagrangian-Fluid}
In the preceding section, we explored the symmetries of fluid mechanics through the lens of gauge theory variables. Here, we will examine these symmetries using the inherent quantities specific to the fluid {in the Lagrangian picture}.
\subsection{Action principle and Noether charge}\label{sec:action-principle}
In this subsection, we start from the first variation of the fluid action \eqref{action-1} and analyze its symmetries and corresponding charges. The variation of \eqref{action-1} yields
\begin{equation}\label{variation-fluid-action}
    \delta S=\rho_{0}\int \dd t \dd^{\,d}y \, \mathcal{E}^{\,i}\delta x_{i}+\rho_{0}\int \dd t \dd^{\,d}y\,  \partial_{\mu}\theta^{\,\mu}\, .
\end{equation}
The equations of motion arise from $\mathcal{E}^{i}=0$.
 The temporal component of the symplectic potential is also given by
\begin{equation}\label{t-com-theta}
    \begin{split}
        \theta^{\,t}=& P_i\, \delta x_i\, , \hspace{1 cm} P_i:= m \dot{x}_i-\frac{\partial V}{\partial \dot{x}_i}\, .
    \end{split}
\end{equation}
The explicit form of $\mathcal{E}^{i}$ and $\theta^{i}$ are given in the appendix \ref{sec:symp-eom}.

The transformation \eqref{inf-diffeos} is the symmetry of the action \eqref{action-1} {in the following sense} {as it leaves the action invariant up to a boundary term}
\begin{equation}\label{variation-action}
    \delta_{\mathcal{Y}}S=\int \dd{t}  \dd^{d}y\, \partial_{i}K^{i}\, , \quad K^{i}:=\mathcal{Y}^{i}L\, ,
\end{equation}
where $L$ denotes the Lagrangian density associated with the action \eqref{action-1}.  {Now we can employ the Noether machinery to read the current related to this symmetry}
\begin{equation}
    J^{t}_{\mathcal{Y}}= P_{i}\,\delta_{\mathcal{Y}} x_i=\mathcal{Y}^{j} P_{i}\partial_{j}x_i\, , \hspace{1 cm}  J_{\mathcal{Y}}^{i}=\theta^{\,i}-K^{i}\, ,
\end{equation}
which is conserved on shell, $\partial_{t}J_{\mathcal{Y}}^{t}+\partial_{i}J_{\mathcal{Y}}^{i}=0$. Once the Noether current is obtained, the Noether charge can be calculated as follows
\begin{equation}\label{charge-fluid}
   Q_{\mathcal{Y}}=\int_{\Sigma_{d}} \dd^d y\, \mathcal{Y}^{j} P_{i}\partial_{j}x_i\, ,
\end{equation}
where $\Sigma_{d}$ is a codimension-one Cauchy surface. Since $\mathcal{Y}^i$ is arbitrary, the expression for the Noether charge represents an infinite number of conserved charges\footnote{One can simply check that the codimension-one charge \eqref{charge-fluid} at the near-equilibrium limit by using the fluid/gauge dictionary \eqref{guage-fluid-dictionary} and volume-preserving condition \eqref{volume-pres-cond-2} reduces to the \eqref{codim-1-charge-p-form} namely, $ Q_{\mathcal{Y}}\to \int_{\Sigma}\star \bold{F}\wedge \dd\bold{\Lambda}\,$.}.
In the subsequent analysis, we will figure out the relationship between these charges and the Kelvin circulation theorem.
\subsection{The Kelvin circulation theorem}\label{sec:Kelvin}
To obtain the Kelvin circulation theorem,  we note that the volume-preserving condition $\partial_{i}\mathcal{Y}^i=0$ yields $\mathcal{Y}^i=\partial_{j}\lambda^{ij}$ \eqref{volume-pres-cond-2}.
Now, we substitute this into the charge expression \eqref{charge-fluid} to get
\begin{align}
Q_{\mathcal{Y}}&=\int_{\Sigma} \dd^d y\, \partial_{k}\lambda^{jk} P_{i}\partial_{j}x_i
=-\int_{\Sigma} \dd^d y\, \lambda^{jk} \partial_{k}\left(P_{i}\partial_{j}x_i\right),%
\end{align}
where the second equality is valid up to a boundary term.
Since this Noether charge is conserved for any anti-symmetric $\lambda^{ij}$, we conclude
$\partial_{t}\partial_{[k}\left(P_{i}\partial_{j]}x_i\right)=0$. This expression is a two-form and one can integrate over a two-dimensional surface $\Sigma_{2}$, namely,
\begin{equation}
    \frac{d}{dt}\int_{\Sigma_2} \dd\Sigma^{kj}\partial_{[k}\left(P_{i}\partial_{j]}x_i\right)=0\, .
\end{equation}
Now using Stokes’ theorem enables us to read it as
\begin{equation}
     \frac{d}{dt} \oint_{\mathcal{C}}\dd y^{j}\, P_{i}\partial_{j}x_i=0\, ,
\end{equation}
where $\mathcal{C}:=\partial \Sigma_2$. 
Alternatively, this can be rewritten as follows
\begin{equation}\label{Kelvin-theorem-fluid}
  \eqbox{  \frac{d}{dt} \Gamma_{\mathcal{C}}=0\, , \hspace{1 cm}  \Gamma_{\mathcal{C}}:=\oint_{\mathcal{C}}\, \vb{P}\cdot \dd\vb{x}\, .}
\end{equation}
Once again, we encounter the Kelvin circulation theorem. However, this time it emerges within the context of fluid dynamics and as a direct consequence of volume-preserving diffeomorphisms.
\subsection{Charge algebra}
In this subsection, we will focus our attention on computing the algebra of charges {in the framework of} fluid mechanics. To read the charge algebra, we introduce the \textit{Noether momentum charge aspect} as follows
\begin{equation}
   q_j:=P_{i}\partial_{j}x_i\, ,
\end{equation}
then the Noether charge \eqref{charge-fluid} is written as $Q_{\mathcal{Y}}=\int_{\Sigma_d} \dd^d y\, \mathcal{Y}^{j} q_{j}$. One can simply show that the variation of the charge aspect is given by
\begin{equation}\label{charge-aspect-var}
    \delta_{\mathcal{Y}}q_{i}=\mathcal{Y}^{j}\partial_{j}q_{i}+q_j \partial_i \mathcal{Y}^j\, .
\end{equation}
We will now calculate the charge algebra using the following equation
$
   \{Q_{\mathcal{Y}}, Q_{\tilde{\mathcal{Y}}}\}:= \delta_{\tilde{\mathcal{Y}}}Q_{\mathcal{Y}}.
$
 By evaluating the right-hand side using equation \eqref{charge-aspect-var}, we arrive at the final result
\begin{equation}\label{charge-algbera-fluid}
     \{q_{i}(\vb{y},t),q_{j}(\vb{y}',t)\}=\delta^{d}(\vb{y}-\vb{y}')(\partial_{j}q_{i}-\partial_{i}q_{j})\, .
\end{equation}
This algebra is a consequence of the Heisenberg algebra for conjugate variables, $\{x_i(\vb{y},t), P_j(\vb{y}',t)\}=\delta_{ij}\delta^{d}(\vb{y}-\vb{y}')$.
One can rephrase the charge algebra \eqref{charge-algbera-fluid} in the following manner
$\{Q_{\mathcal{Y}},Q_{\tilde{\mathcal{Y}}}\}=Q_{[\mathcal{Y},\tilde{\mathcal{Y}}]}$.
This demonstrates the representation theorem in the covariant phase space method, which asserts that the charge algebra is isomorphic to the symmetry algebra up to the central extension terms {(see \cite{Compere:2018aar} for a review)}. In our case,  there are no central terms.
\section{Outlook}\label{sec:discussion}

\begin{table}[t]
\begin{tabular}{ |c|c|c|  }
\hline
{Lagrangian Fluid }& {$(d-1)$-form gauge theory} \\
\hline
 Velocity  &  Electric field\\
 Density & Magnetic field\\
{Vortex} & {$(d-2)$-brane}\\
{Circulation} & {Electric charge} \\
{Relabeling freedom of fluid parcels}  &  {Gauge symmetry}\\
{Kelvin's circulation theorem}  & {Gauss's law}\\
{Mass continuity equation}  & {Bianchi identity}\\
 \hline
\end{tabular}
\caption{fluid/$(p+1)$-form dictionary}\label{dictionary}
\end{table}

In this study, we delved into an inviscid $d+1$ dimensional Lagrangian fluid. We probed the inherent volume-preserving diffeomorphisms of the fluid and derived the associated conserved charges. The Kelvin circulation theorem was interpreted as the global charge among these charges. Furthermore, we uncovered the non-Abelian algebra of volume-preserving diffeomorphisms and the corresponding charges. 

{Moreover, we examined the fluid's near-equilibrium state. A key finding of this study is that a $(d-1)$-form gauge theory effectively describes fluids in this regime. The gauge theory has Abelian gauge symmetry, which is well-aligned with linearized volume-preserving diffeomorphisms in the original Lagrangian fluid.}
{Notably,} we found the Kelvin circulation theorem on the fluid side corresponds to the Gauss law on the gauge theory side. Lastly, we formulated a fluid/$(d-1)$-form dictionary, which is presented in Table \ref{dictionary}.

We conclude with a few remarks and potential directions for future research.

We first note the differences between the codimension-one charge in the fluid side \eqref{charge-fluid} and the codimension-two charge in the gauge theory side \eqref{codim-2-charge}. Our symmetry generators in the fluid side $\mathcal{Y}(\vb{y})$, have arbitrary spatial coordinate dependence, so one may want to compare them
with gauge symmetries and anticipate a codimension-two expression for the corresponding charges. However, as evident from \eqref{charge-fluid}, the corresponding Noether charges are codimension-one. The key point here is that gauge symmetries involve arbitrary functions of all \textit{spacetime} coordinates, while in the fluid case, we have diffeomorphisms that depend only on \emph{spatial} coordinates (not time). In this regard, it is sometimes said that diffeomorphisms in fluid mechanics are \textit{global} symmetries, and it is natural to obtain a codimension-one expression for the Noether charge associated with volume-preserving diffeomorphisms.

For our second comment, it’s important to note that when two theories are dual, the number of degrees of freedom on both sides must match. On the fluid side, we have $d$ scalar fields {$x_{i}(y)$} along with $d$ volume-preserving diffeomorphisms. The volume-preserving condition reduces one of the arbitrary functions in diffeomorphisms leaving us with just one degree of freedom for the fluid system, namely $d-(d-1)=1$. On the gauge theory side, we can calculate the number of propagating degrees of freedom by considering a $(p+1)$-form theory in $d+1$ spacetime dimensions. It can be {proven} that the number of propagating degrees of freedom in such a theory is given by the binomial coefficient $C(d-1,p+1)$. For example, in Maxwell’s theory $(p=0)$ with $d+1$ dimensions, there are $d-1$ degrees of freedom. In our case, this results in $C(d-1,d-1)=1$ degree of freedom, which aligns with our findings on the fluid side.

Our study was conducted within the framework of the Lagrangian picture. It would be intriguing to express the $d+1$ dimensional fluid in the language of gauge theory in the Eulerian picture. This has already been accomplished for incompressible Eulerian fluid in $1+2$ and $1+3$ spacetime dimensions in \cite{Eling:2023apf, Eling:2023iyx}. 
{As a comparison with our work, we note that the gauge structure of both is the same. To put it another way, in both pictures, we are dealing with $(d-1)$-form gauge theories (at least for $d=2,3$). This is not surprising and can be anticipated because the rank of the gauge field is dictated by the Kelvin theorem that holds true regardless of whether fluid is described in the Lagrangian or Eulerian framework.}

In reference \cite{Sheikh-Jabbari:2023eba}, the memory effect, as a direct consequence of the gauge theory formulation, has been examined in detail. It would be worthwhile to consider the implications of the gauge theory formulation within the Lagrangian picture.

Our formulation is currently limited to inviscid fluids. Therefore, considering the effects of dissipation presents another interesting avenue for future research.\\

\textbf{Acknowledgements.}
 {We are grateful to M.M. Sheikh-Jabbari for bringing \cite{Susskind:2001fb} to our attention and for his insightful comments and discussion. We would also like to thank A. Najafi for the fruitful discussion.}

\appendix
\section{Fluid equation of motion and symplectic potential}\label{sec:symp-eom}
The first variation of the action \eqref{action-1} yields \eqref{variation-fluid-action} where the explicit form of the equation of motion and the spatial components of symplectic potential is given by
\begin{equation}\label{eom-fluid}
    \mathcal{E}_{i}=-m \Ddot{x}_{i}-\frac{\partial V}{\partial x_i}+\frac{d}{dt}\left(\frac{\partial V}{\partial \dot{x}_i}\right)-\partial_{j}\left(\rho\frac{\partial V}{\partial \rho}\left(\frac{\partial x_{i}}{\partial y_{j}}\right)^{-1}\right)=0\, ,
\end{equation}
\begin{equation}
    \theta^{i}=\rho \frac{\partial V}{ \partial \rho} \left(\frac{\partial x_j}{\partial y_i}\right)^{-1} \delta x_{j}\, .
\end{equation}
\bibliography{reference}

\begin{thebibliography}{25}
\expandafter\ifx\csname natexlab\endcsname\relax\def\natexlab#1{#1}\fi
\expandafter\ifx\csname bibnamefont\endcsname\relax
  \def\bibnamefont#1{#1}\fi
\expandafter\ifx\csname bibfnamefont\endcsname\relax
  \def\bibfnamefont#1{#1}\fi
\expandafter\ifx\csname citenamefont\endcsname\relax
  \def\citenamefont#1{#1}\fi
\expandafter\ifx\csname url\endcsname\relax
  \def\url#1{\texttt{#1}}\fi
\expandafter\ifx\csname urlprefix\endcsname\relax\def\urlprefix{URL }\fi
\providecommand{\bibinfo}[2]{#2}
\providecommand{\eprint}[2][]{\url{#2}}

\bibitem[{\citenamefont{Morrison}(1998)}]{RevModPhys.70.467}
\bibinfo{author}{\bibfnamefont{P.~J.} \bibnamefont{Morrison}},
  \bibinfo{journal}{Rev. Mod. Phys.} \textbf{\bibinfo{volume}{70}},
  \bibinfo{pages}{467} (\bibinfo{year}{1998}),
  \urlprefix\url{https://link.aps.org/doi/10.1103/RevModPhys.70.467}.

\bibitem[{\citenamefont{{Shepherd}}(1990)}]{1990AdGeo..32..287S}
\bibinfo{author}{\bibfnamefont{T.~G.} \bibnamefont{{Shepherd}}},
  \bibinfo{journal}{Advances in Geophysics} \textbf{\bibinfo{volume}{32}},
  \bibinfo{pages}{287} (\bibinfo{year}{1990}).

\bibitem[{\citenamefont{Salmon}(1988)}]{Salmon1988HAMILTONIANFM}
\bibinfo{author}{\bibfnamefont{R.}~\bibnamefont{Salmon}},
  \bibinfo{journal}{Annual Review of Fluid Mechanics}
  \textbf{\bibinfo{volume}{20}}, \bibinfo{pages}{225} (\bibinfo{year}{1988}).

\bibitem[{\citenamefont{Vladimir I.~Arnold}(2008)}]{V.Arnold}
\bibinfo{author}{\bibfnamefont{B.~A.~K.} \bibnamefont{Vladimir I.~Arnold}},
  \bibinfo{journal}{Annual Review of Fluid Mechanics}
  \textbf{\bibinfo{volume}{125}} (\bibinfo{year}{2008}).

\bibitem[{\citenamefont{Henneaux and Teitelboim}(1986)}]{Henneaux:1986ht}
\bibinfo{author}{\bibfnamefont{M.}~\bibnamefont{Henneaux}} \bibnamefont{and}
  \bibinfo{author}{\bibfnamefont{C.}~\bibnamefont{Teitelboim}},
  \bibinfo{journal}{Found. Phys.} \textbf{\bibinfo{volume}{16}},
  \bibinfo{pages}{593} (\bibinfo{year}{1986}).

\bibitem[{\citenamefont{Delplace et~al.}(2017)\citenamefont{Delplace, Marston,
  and Venaille}}]{Delplace_2017}
\bibinfo{author}{\bibfnamefont{P.}~\bibnamefont{Delplace}},
  \bibinfo{author}{\bibfnamefont{J.~B.} \bibnamefont{Marston}},
  \bibnamefont{and} \bibinfo{author}{\bibfnamefont{A.}~\bibnamefont{Venaille}},
  \bibinfo{journal}{Science} \textbf{\bibinfo{volume}{358}},
  \bibinfo{pages}{1075} (\bibinfo{year}{2017}),
  \urlprefix\url{https://doi.org/10.1126%2Fscience.aan8819}.

\bibitem[{\citenamefont{Venaille and Delplace}(2021)}]{Venaille_2021}
\bibinfo{author}{\bibfnamefont{A.}~\bibnamefont{Venaille}} \bibnamefont{and}
  \bibinfo{author}{\bibfnamefont{P.}~\bibnamefont{Delplace}},
  \bibinfo{journal}{Physical Review Research} \textbf{\bibinfo{volume}{3}}
  (\bibinfo{year}{2021}),
  \urlprefix\url{https://doi.org/10.1103%2Fphysrevresearch.3.043002}.

\bibitem[{\citenamefont{Tong}(2022)}]{Tong:2022gpg}
\bibinfo{author}{\bibfnamefont{D.}~\bibnamefont{Tong}} (\bibinfo{year}{2022}),
  \eprint{2209.10574}.

\bibitem[{\citenamefont{Sheikh-Jabbari
  et~al.}(2023)\citenamefont{Sheikh-Jabbari, Taghiloo, and
  Vahidinia}}]{Sheikh-Jabbari:2023eba}
\bibinfo{author}{\bibfnamefont{M.~M.} \bibnamefont{Sheikh-Jabbari}},
  \bibinfo{author}{\bibfnamefont{V.}~\bibnamefont{Taghiloo}}, \bibnamefont{and}
  \bibinfo{author}{\bibfnamefont{M.~H.} \bibnamefont{Vahidinia}}
  (\bibinfo{year}{2023}), \eprint{2302.04912}.

\bibitem[{\citenamefont{Eling}(2023{\natexlab{a}})}]{Eling:2023iyx}
\bibinfo{author}{\bibfnamefont{C.}~\bibnamefont{Eling}}
  (\bibinfo{year}{2023}{\natexlab{a}}), \eprint{2305.04394}.

\bibitem[{\citenamefont{Eling}(2023{\natexlab{b}})}]{Eling:2023apf}
\bibinfo{author}{\bibfnamefont{C.}~\bibnamefont{Eling}}
  (\bibinfo{year}{2023}{\natexlab{b}}), \eprint{2310.12475}.

\bibitem[{\citenamefont{Nastase and Sonnenschein}(2023)}]{Nastase:2023rou}
\bibinfo{author}{\bibfnamefont{H.}~\bibnamefont{Nastase}} \bibnamefont{and}
  \bibinfo{author}{\bibfnamefont{J.}~\bibnamefont{Sonnenschein}}
  (\bibinfo{year}{2023}), \eprint{2303.15229}.

\bibitem[{\citenamefont{Dayi}(2023)}]{Dayi:2023ckd}
\bibinfo{author}{\bibfnamefont{O.~F.} \bibnamefont{Dayi}}
  (\bibinfo{year}{2023}), \eprint{2310.13517}.

\bibitem[{\citenamefont{Bahcall and Susskind}(1991)}]{Bahcall:1991an}
\bibinfo{author}{\bibfnamefont{S.}~\bibnamefont{Bahcall}} \bibnamefont{and}
  \bibinfo{author}{\bibfnamefont{L.}~\bibnamefont{Susskind}},
  \bibinfo{journal}{Int. J. Mod. Phys. B} \textbf{\bibinfo{volume}{5}},
  \bibinfo{pages}{2735} (\bibinfo{year}{1991}).

\bibitem[{\citenamefont{Susskind}(2001)}]{Susskind:2001fb}
\bibinfo{author}{\bibfnamefont{L.}~\bibnamefont{Susskind}}
  (\bibinfo{year}{2001}), \eprint{hep-th/0101029}.

\bibitem[{\citenamefont{Afshar et~al.}(2018)\citenamefont{Afshar, Esmaeili, and
  Sheikh-Jabbari}}]{Afshar:2018apx}
\bibinfo{author}{\bibfnamefont{H.}~\bibnamefont{Afshar}},
  \bibinfo{author}{\bibfnamefont{E.}~\bibnamefont{Esmaeili}}, \bibnamefont{and}
  \bibinfo{author}{\bibfnamefont{M.~M.} \bibnamefont{Sheikh-Jabbari}},
  \bibinfo{journal}{JHEP} \textbf{\bibinfo{volume}{05}}, \bibinfo{pages}{042}
  (\bibinfo{year}{2018}), \eprint{1801.07752}.

\bibitem[{\citenamefont{Compere}(2007)}]{Compere:2007vx}
\bibinfo{author}{\bibfnamefont{G.}~\bibnamefont{Compere}},
  \bibinfo{journal}{Phys. Rev. D} \textbf{\bibinfo{volume}{75}},
  \bibinfo{pages}{124020} (\bibinfo{year}{2007}), \eprint{hep-th/0703004}.

\bibitem[{\citenamefont{Esmaeili}(2020)}]{Esmaeili:2020eua}
\bibinfo{author}{\bibfnamefont{E.}~\bibnamefont{Esmaeili}}, Ph.D. thesis,
  \bibinfo{school}{IPM, Tehran} (\bibinfo{year}{2020}), \eprint{2010.13922}.

\bibitem[{\citenamefont{Gomes}(2023)}]{Gomes:2023ahz}
\bibinfo{author}{\bibfnamefont{P.~R.~S.} \bibnamefont{Gomes}},
  \bibinfo{journal}{SciPost Phys. Lect. Notes} \textbf{\bibinfo{volume}{74}},
  \bibinfo{pages}{1} (\bibinfo{year}{2023}), \eprint{2303.01817}.

\bibitem[{\citenamefont{Cordova et~al.}(2022)\citenamefont{Cordova, Dumitrescu,
  Intriligator, and Shao}}]{Cordova:2022ruw}
\bibinfo{author}{\bibfnamefont{C.}~\bibnamefont{Cordova}},
  \bibinfo{author}{\bibfnamefont{T.~T.} \bibnamefont{Dumitrescu}},
  \bibinfo{author}{\bibfnamefont{K.}~\bibnamefont{Intriligator}},
  \bibnamefont{and} \bibinfo{author}{\bibfnamefont{S.-H.} \bibnamefont{Shao}},
  in \emph{\bibinfo{booktitle}{{Snowmass 2021}}} (\bibinfo{year}{2022}),
  \eprint{2205.09545}.

\bibitem[{\citenamefont{Schafer-Nameki}(2023)}]{Schafer-Nameki:2023jdn}
\bibinfo{author}{\bibfnamefont{S.}~\bibnamefont{Schafer-Nameki}}
  (\bibinfo{year}{2023}), \eprint{2305.18296}.

\bibitem[{\citenamefont{Brennan and Hong}(2023)}]{Brennan:2023mmt}
\bibinfo{author}{\bibfnamefont{T.~D.} \bibnamefont{Brennan}} \bibnamefont{and}
  \bibinfo{author}{\bibfnamefont{S.}~\bibnamefont{Hong}}
  (\bibinfo{year}{2023}), \eprint{2306.00912}.

\bibitem[{\citenamefont{Sharpe}(2015)}]{Sharpe:2015mja}
\bibinfo{author}{\bibfnamefont{E.}~\bibnamefont{Sharpe}},
  \bibinfo{journal}{Fortsch. Phys.} \textbf{\bibinfo{volume}{63}},
  \bibinfo{pages}{659} (\bibinfo{year}{2015}), \eprint{1508.04770}.

\bibitem[{\citenamefont{McGreevy}(2022)}]{McGreevy:2022oyu}
\bibinfo{author}{\bibfnamefont{J.}~\bibnamefont{McGreevy}}
  (\bibinfo{year}{2022}), \eprint{2204.03045}.

\bibitem[{\citenamefont{Compère and Fiorucci}(2019)}]{Compere:2018aar}
\bibinfo{author}{\bibfnamefont{G.}~\bibnamefont{Compère}} \bibnamefont{and}
  \bibinfo{author}{\bibfnamefont{A.}~\bibnamefont{Fiorucci}},
  \bibinfo{journal}{Lect. Notes Phys.} \textbf{\bibinfo{volume}{952}},
  \bibinfo{pages}{150} (\bibinfo{year}{2019}), \eprint{1801.07064}.

\end{thebibliography}

\end{document}